\documentclass[prb,superscriptaddress,groupaddress,letterpaper,twocolumn,pacs]{revtex4}
\usepackage{natbib}
\usepackage{cancel}
\usepackage{ulem}
\usepackage{amsthm}
\usepackage{amsmath}
\usepackage{amssymb}
\usepackage{amsfonts}
\usepackage{graphicx}
\usepackage{verbatim}
\usepackage{color}
\usepackage{float}
\usepackage{longtable}

\definecolor{dark-red}{rgb}{0.45,0.0,0.0}
\definecolor{dark-green}{rgb}{0.0,0.45,0.0}
\definecolor{dark-blue}{rgb}{0.0,0.0,0.85}

\begin{document}

\preprint{}
\title{Sailing through the big crunch-big bang transition }
\author{Itzhak Bars}
\affiliation{Department of Physics and Astronomy, University of Southern
California, Los
Angeles, CA, 90089-0484, USA,}
\author{Paul Steinhardt}
\affiliation{Physics Department, Princeton University, Princeton NJ08544,
USA,}
\author{Neil Turok}
\affiliation{Perimeter Institute for Theoretical Physics, Waterloo, ON
N2L 2Y5, Canada.}
\date{November 20, 2013}

\begin{abstract}
In a recent series of papers, we have shown that theories with scalar
fields
coupled to gravity (\textit{e.g.}, the standard model) can be lifted to a
Weyl-invariant equivalent theory in which it is possible to unambiguously
trace the classical cosmological evolution through the transition from
big
crunch to big bang. The key was identifying a sufficient number of
finite,
Weyl-invariant conserved quantities to uniquely match the fundamental
cosmological degrees of freedom across the transition. In so doing we had
to
account for the well-known fact that many Weyl-invariant quantities
diverge
at the crunch and bang. Recently, some authors rediscovered a few of
these
divergences and concluded based on their existence alone that the
theories
cannot be geodesically complete. In this note, we show that this
conclusion
is invalid. Using conserved quantities we explicitly construct the
complete
set of geodesics and show that they pass continuously through the big
crunch-big bang transition.
\end{abstract}

\maketitle

In the standard big bang inflationary model \cite{Guth}, the cosmic
singularity problem is left unresolved and the cosmology is geodesically
incomplete. Consequently, the origin of space and time and the peculiar,
exponentially fine-tuned initial conditions required to begin inflation
\cite%
{Penrose,GibbonsTurok} are not explained. In a recent series of papers
\cite%
{inflationBC,cyclicBCT,cyclic-Bars,Bars:2011aa,Bars:2012mt,Bars:2012fq,NewestBST,BSTHiggs}%
, we have shown how to construct the complete set of homogeneous
classical cosmological solutions of the standard model coupled to
gravity, in which the cosmic singularity is replaced by a bounce:
the smooth transition from contraction and big crunch to big bang
and expansion. These are \textit{generic} geodesically complete
solutions that can, for
example, naturally incorporate the cyclic theory of the universe \cite%
{cyclic1,cyclic2} in which it is proposed that large-scale smoothness,
flatness and nearly scale-invariant perturbations are generated during
the
periods of slow contraction preceding each big bang.

The key to our construction of classical geodesically complete solutions
was
to \textquotedblleft lift\textquotedblright\ the action (\textit{e.g.},
the
standard model coupled to Einstein gravity) to a Weyl-invariant
equivalent
theory. We then identified a number of Weyl-invariant finite quantities
\cite%
{Bars:2011aa} that are conserved near cosmological singularities for
symmetry reasons. Our proposal was to match these quantities across the
singularities which separate the patches of spacetime describing the big
crunch-big bang transition. We showed there were sufficiently many such
conserved quantities to ensure a unique match for all cosmological
fields.
In so doing, we necessarily had to pay attention to the well-known fact
\cite%
{c1,c2,c3,c4} that many Weyl-invariant quantities diverge at the bounce,
such as the Weyl curvature, $C_{\nu \lambda \delta }^{\mu }$. Recently,
Carrasco, Chemisanny and Kallosh \cite{c} and Kallosh and Linde \cite{a2}
rediscovered some of these divergences and, without paying attention to
our
discussion of conserved Weyl-invariant finite quantities, claimed the
divergences necessarily spoil the geodesic completeness of our proposed
big
crunch-big bang transition. In this note, we demonstrate that this naive
claim is incorrect.

To be sure, what is presented here is a straightforward elaboration of
what
was already proven in our earlier papers \cite%
{inflationBC,cyclicBCT,cyclic-Bars,Bars:2011aa,Bars:2012mt,Bars:2012fq,NewestBST,BSTHiggs}%
. Once we identified a sufficient number of conserved finite Weyl-
invariant
quantities to determine a unique continuation of all the fundamental
cosmological fields (\textit{e.g.}, scalar fields and metric) across all
patches of field space, it should be obvious that the spacetime is
geodesically complete because the geodesics of particles in the theory
are
all expressed in terms of these cosmological fields, as detailed below.
It
remains true that there exist infinitely many Weyl-invariant quantities
that
diverge at the crunch or bang, but these are irrelevant to the geodesic
completeness. In fact, even for these quantities, the field continuation
proposed in our papers uniquely determines their evolution before and
after
they go singular.

To illustrate the point, we focus on the vicinity of the big crunch or
big
bang where it suffices to consider a simplified standard model with a
Higgs-like scalar field conformally coupled to gravity plus radiation.
Following the procedure in \cite%
{inflationBC,cyclicBCT,cyclic-Bars,Bars:2011aa,Bars:2012mt,Bars:2012fq,NewestBST,BSTHiggs}%
, the \textquotedblleft lifted\textquotedblright\ action of the
standard model is achieved by adding an extra scalar field $\phi $
and imposing Weyl symmetry so the number of gauge invariant physical
degrees of freedom remain the same. What is achieved in this way is
the inclusion of all patches of the fields that leads to a
geodesically complete cosmology as explained in more detail in
[\onlinecite{NewestBST,BSTHiggs}]. The leading contributions near
the big crunch or the big bang are \cite{Bars:2011aa}:
\begin{equation}
\int d^{4}x\sqrt{-g}\left[ \frac{1}{2}\left( (\partial \phi )^{2}-
(\partial
h)^{2}\right) +\frac{1}{12}(\phi ^{2}-h^{2})R\right] .
\label{conformal action}
\end{equation}%
where $h^{2}\equiv H^{\dagger }H$ for Higgs doublet $H$. The
complete description also includes a term describing the radiation
(see below). All other contributions to the action, including matter
fields, as well as density perturbations  become negligible in this
limit and the cosmic evolution becomes \textit{smoothly ultralocal},
meaning that spatial gradients become dynamically negligible
\cite{Rendall}. The fact that the spacetime may be treated as
spatially homogeneous near the singularity also allows us to find
all of its geodesics.

The action is invariant under the local gauge transformations
$g_{\mu \nu} \rightarrow \Omega^{-2}(x^{\mu}) g_{\mu \nu}$, $\phi
\rightarrow \Omega(x^{\mu}) \phi$ and $h \rightarrow \Omega(x^{\mu})
h$. Although the lift introduces a second scalar field $\phi$ with
wrong-sign kinetic energy, it is obviously a gauge-artifact since
fixing a Weyl gauge
in which $\phi_E \equiv (\sqrt{6}/\kappa)\cosh{\kappa \sigma/\sqrt{6}}$
and $%
h_E \equiv (\sqrt{6}/\kappa) \sinh{\kappa \sigma/\sqrt{6}}$, where ($%
\kappa^2\equiv 8 \pi G$, with $G$ Newton's constant) converts the
action to Einstein gravity plus canonical scalar field $\sigma$ with
no ghost degrees of freedom in the incoming and outgoing
cosmological states. Using the Bianchi I, VIII or IX metrics
including anisotropy, the line element near the big crunch and bang
is:\cite{Bars:2011aa}
\begin{equation}  \label{Kasmetric}
\begin{split}
ds^{2} = & a^{2}\left( \tau\right) \Big[ -d\tau^{2}+ e^{-\sqrt{8/3}%
\kappa\alpha_{1}} d\sigma_3^{2} + \\
& e^{\sqrt{2/3}\kappa\alpha_{1}}\left( e^{\sqrt{2}\kappa \alpha_{2}}
d\sigma_1^{2} +e^{-\sqrt{2}\kappa\alpha_{2}} d\sigma_2^{2}\right) \Big]
\end{split}%
\end{equation}
where $\tau$ is the conformal time and $\alpha_{1,2}(\tau)$ parameterize
the
anisotropy. The $d\sigma_{1,2,3}$ generically include information about
the
spatial curvature; however, since the spatial curvature is negligible
near a
big crunch or big bang, $d \sigma_{1,2,3}$ reduce locally to
$dx_{1,2,3}$,
respectively, resulting in the Kasner-type metric.

Another useful gauge choice, dubbed $\gamma$-gauge fixes
$g_{\gamma}=-1$ or equivalently the scale factor $a_{\gamma}=1$, in
which case the action specialized to homogeneous fields is:
\begin{eqnarray}  \label{effect}
\int d\tau\left( \frac{1}{2e}\left[ -\dot{\phi}_{\gamma}^2 + \dot{h}%
_{\gamma}^2 + \frac{\kappa^2}{6}(\phi_{\gamma}^2 -
h_{\gamma}^2)(\dot{\alpha}%
_1^2+ \dot{\alpha}_2^2)\right] -e \rho_r \right),
\end{eqnarray}
where $e(\tau)$ is the lapse function and the radiation density is $%
\rho_r/a^4(\tau)$ where $\rho_r$ is constant.

In this gauge, it is straightforward to find the complete set of
solutions that continuously track the evolution of $\phi _{\gamma
}$, $h_{\gamma }$ and $\alpha _{1,2}$ through a big crunch, a brief
interlude of antigravity, and then a big bang. Expressing the
solution in terms of Einstein gauge fields (indicated by the
subscript $E$) we obtain \cite{Bars:2011aa}:
\begin{eqnarray}
a_{E}^{2}\left( \tau \right) &=&2\left\vert \tau \right\vert \left\vert
p+\rho _{r}\tau \right\vert \\
\alpha _{1,2}\left( \tau \right) &=&\frac{p_{1,2}}{2p}\ln \left\vert
\frac{%
\tau }{T_{1,2}\left( p+\rho _{r}\tau \right) }\right\vert \\
a_{E}^{2}h_{E}^{2} &=&\frac{1}{2}|\tau ||p+\rho _{r}\tau |\times \\
&&\left( |T_{3}(\rho _{r}+\frac{p}{\tau })|^{p_{3}/2p}-|T_{3}(\rho _{r}+%
\frac{p}{\tau })|^{-p_{3}/2p}\right) ^{2}  \notag
\end{eqnarray}%
where $\tau $ is conformal time in units where $\kappa =\sqrt{6}$ and $%
T_{1,2,3}$ are integration constants. Note that the solution for
$a^{2}h^{2}=a_{E}^{2}h_{E}^{2}$ given above is Weyl invariant. The
crunch occurs at $\tau _{c}=-p/\rho _{r}$ and the bang at $\tau =0$
with the period of antigravity in between. The constants $p_{1,2,3}$
are the finite \textit{conserved} values of the canonical momenta of
the \textit{fields} at the crunch or
bang: $\pi _{3}=a_{E}^{2}\dot{h}_{E}\rightarrow p_{3},$ $\pi
_{1,2}=a_{E}^{2}%
\dot{\alpha}_{1,2}\rightarrow p_{1,2},$ and $p\equiv \sqrt{%
p_{1}^{2}+p_{2}^{2}+p_{3}^{2}}$. A $|p_3|$ which is larger than
$\sqrt{15}(p_1^2+p_2^2)^{1/2}$ insures the avoidance of the
mixmaster behavior \cite{Misner,BKL} even when spatial curvature is
present \cite{cyclic-Bars}. In Ref.~[\onlinecite{Bars:2011aa}], we
describe a total of 15 conserved Noether charges that are finite at
the crunch or bang and whose conservation across the singularities
uniquely determines the solutions given above.

\begin{center}
\includegraphics[height=2in] {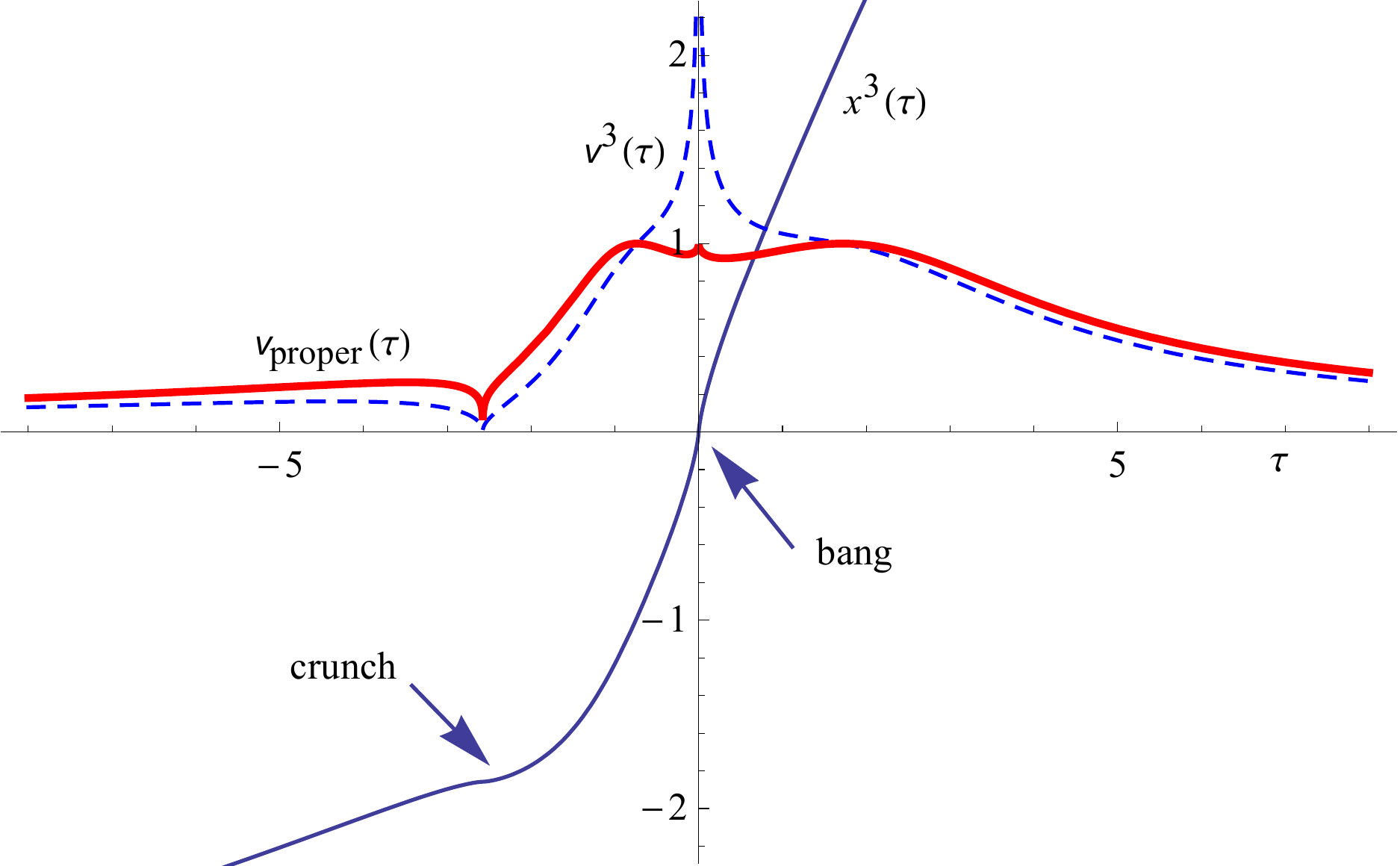} \label{t1.eps}
\end{center}

{\footnotesize Figure 1. A typical massive particle geodesic,
showing the continuous passage through the big crunch and big bang
singularities. In this example, the coordinate velocity
$v^3=\dot{x}^{3}$ goes to zero at the crunch and infinity at the
bang. The proper speed of the particle with respect to particles
comoving with the $x^i$ coordinates, $v_{proper}=\sqrt{g_{3ij}%
\dot{x}^{i}\dot{x}^{j}}/\sqrt{g_{3ij}\dot{x}^{i}\dot{x}^{j}+m^{2}a^{2}}\
$, never exceeds unity and touches zero at the crunch and unity at
the bang. Likewise, the coordinate $x^{3}$ is finite and continuous
throughout. The numerical values of the parameters used to generate
this
plot are $p_1=-1/4$,\ $p_2=0$,\ $p_3=1$,\ $T_1=1$,\ $T_2=1$,\ $T_3=1$,\
$%
k_1=0$,\ $k_2=0$,\ $k_3=1$,\ $g_p=1$,\ $\rho_r=0.4$.}\medskip

We now consider the geodesics of massive particles in the standard model
in
this geometry. The \textit{Weyl-invariant} action for a particle moving
in a
gravitational background can be expressed as:
\begin{equation}\label{geoaction}
\mathcal{S}_{\text{particle}}=-\int d\lambda ~m\left( x\right) \sqrt{-
\dot{x}^{\mu }%
\dot{x}^{\nu }g_{\mu \nu }\left( x\right) },
\end{equation}%
where $x^{\mu }(\lambda )$ is a function of the affine parameter $\lambda
$.
Note that $m\left( x\right)$ is generally $x$-dependent in theories like
the
standard model when the Higgs field contributes to the mass of particles:
$%
m\left( x\right) =g_p h\left( x\right) $, where $g_p$ is a
dimensionless coupling of the particle field to the Higgs field.
From this action, an explicit expression for \textit{all} geodesics
in an anisotropic Kasner universe can be derived, exploiting the
spatial homogeneity of the metric \cite{cyclicBCT,NewestBST}:
\begin{equation} \label{geosol}
x^{i}\left( \tau \right) =q^{i}+\int^{\tau }d\tau ^{\prime }\frac{%
g_{3}^{ij}\left( \tau ^{\prime }\right)
k_{j}}{\sqrt{g_{3}^{kl}\left( \tau ^{\prime }\right)
k_{k}k_{l}+m^{2}\left( \tau ^{\prime }\right) a^{2}\left( \tau
^{\prime }\right) }  }
\end{equation}%
where $k_{i}$ are the spatial components of the \textit{conserved}
particle
momentum, $q_{i}$ is the initial position, and $g_{3}^{ij}\left( \alpha
_{1}\left( \tau \right) ,\alpha _{2}\left( \tau \right) \right) $ is the
inverse of the Kasner space-space metric appearing inside the square
brackets in Eq.~(\ref{Kasmetric}).

With this expression, Eq.~(\ref{geosol}), and noting that the
integral converges for all physical parameters of the fields that
determine the spacetime ($p_{1,2,3}, T_{1,2,3},g_p,\rho_r$) and all
parameters of the geodesics ($k_{1,2,3},q^{1,2,3}$), we are
effectively done with the proof of geodesic completeness. We stress
that our complete solutions for
$%
\alpha _{1,2}$ and the gauge invariant combination $a(\tau)m(\tau
)=g_p a_{E}(\tau )h_{E}(\tau )=g_p h_{\gamma }(\tau )$ are
continuous and sufficiently well-behaved at the crunch and bang;
this insures that the geodesics constructed from them are also
continuous. The fact that the Weyl curvature and the Weyl-invariant
quantities discussed in Refs.~[\onlinecite{c,a2}] diverge does not
affect this conclusion one iota. This is clearly exemplified in Fig.
1 where we have chosen a typical geodesic and computed the behavior
of the geodesic component $x^3(\tau)$ and the proper speed. Both are
continuous throughout the big crunch-big bang transition and the
proper speed is bounded by the speed of light, all despite the fact
that the Weyl curvature and other quantities diverge and the
Weyl-invariant metric is singular. Even though we do not expect to
find a coordinate system to remove all curvature singularities,
these are not sufficiently severe in our case to prevent the
geodesics from completing their cosmic journeys.

We emphasize that this note pertains to \textit{all} the solutions
of the field equations in the vicinity of the cosmological
singularities and \textit{all} the geodesics in those geometries, as
obtained using our proposal.\cite{added} This includes both
\textit{massive} and \textit{massless} particles.\cite{added} (In
our general geodesic expression Eq.(\ref{geosol}) that is expressed
in terms of momenta, all that is needed to cover the case of
lightlike geodesics  is to set the mass or Higgs coupling $g_p$ to
zero. As is well known, this can also be obtained from the particle
action in Eq.(\ref{geoaction}) by first defining the canonical
momenta and keeping the momenta fixed while taking the zero mass
limit; or, equivalently, rewriting the action Eq.(\ref{geoaction})
in the first order formalism and then taking the limit.)

Our central point is that the continuation of the geometry beyond
the singularity is established because we have shown in our case
that all geodesics go through the relevant singularities.
Classically, this geodesic completeness is not affected by the
divergent curvatures that we \cite{seeOld} and others have
identified.\cite{added} In fact, the  completion of the geometry is
not supposed to eliminate the curvature singularities. Rather, it is
supposed to show that, despite the curvature singularities, physical
information can and does journey from cycle to cycle through the
cosmological singularities \cite{added}.
 Hence, we can claim the geodesic completeness for
all \textit{homogeneous} cosmological field configurations of the
standard model coupled to gravity.
Further details and a more
thorough discussion of geodesics, geodesic deviation and geodesic
completeness in Weyl-invariant theories are given in Ref.~[%
\onlinecite{nextgeo}].

Of course, our purely classical analysis does not include strong quantum gravity effects near the singularities because the technology does not yet exist to do those computations.\cite{added}  Nevertheless, finding
classical geodesical completeness and the complete set of classical
solutions is very useful.  Having the guidance and physical insight of classical analysis is
often a reasonable starting point in understanding quantum theoretic
descriptions of physical phenomena, especially when there are indications of new physics, as presented here.  For example, suppose one wished to
pursue physics near
the singularity in the framework of string theory. Our solutions
provide the starting point because they provide  geodesically complete cyclic
background geometries (metric and dilaton) consistent with perturbative
worldsheet conformal symmetry as required by the quantization of the
string moving in backgrounds. Our classical calculations suggest exciting new phenomena to be explored. Based on historical precedents, it is reasonable to suppose that some or all qualitative features will survive quantum (string) corrections.

We thank A. Ijjas for comments on the manuscript. Research at Perimeter
Institute is supported by the Government of Canada through Industry
Canada
and by the Province of Ontario through the Ministry of Research and
Innovation. This research was partially supported by the U.S. Department
of
Energy under grant number DE-FG03-84ER40168 (IB) and under grant number
DE-FG02-91ER40671 (PJS).

\end{document}